# Proximity Induced Superconductivity and Multiple Andreev Reflections in Few-Layer-Graphene

A.Shailos[1], W. Nativel[2], A. Kasumov[1], C. Collet[2], M. Ferrier[1], S. Guéron[1],

R. Deblock[1] and H. Bouchiat[1]

[1]Univ. Paris-Sud, CNRS, UMR 8502, F-91405 Orsay Cedex, France
[2]Thales Research and Technology, F-91767 Palaiseau Cedex, France

A. Shailos: shailos@lps.u-psud.fr

ABSTRACT

We have investigated electronic transport of few-layer-graphene (FLG) connected to superconducting electrodes. The device is prepared by mechanical exfoliation of graphite. A small mesa of FLG is placed on the surface of an insulating Alumina layer over silicon substrate, and is connected with two tungsten electrodes, separated by 2.5 µm, grown by focused ion beam. While tungsten electrodes are superconducting below 4 K, proximity induced superconductivity in FLG is observed below 1 K with a large differential resistance drop at low bias. Signatures of multiple Andreev reflections are observed at peaks located at voltages corresponding to sub-multiple values of $2\Delta/e$ where $\Delta$ is the superconducting gap of the electrodes.

## I. Introduction

Transport properties of the graphene plane have recently received a lot of attention on both experimental and theoretical point of view [1-5]. On one hand, this is due to its 2D character and on the other hand, to its remarkable band structure combining a linear dispersion relation of electronic wave functions similar to mass-less particles and a perfect electron hole symmetry. The Fermi surface consists of two cones touching at one singular, so called Dirac point, where the density of states is zero. In particular the physics at high magnetic field in the quantum Hall effect regime has been shown to exhibit spectacular signatures of this special band structure [4, 5]. The physics in low magnetic field presents also quite surprising features. In particular it was pointed out that weak localization which is the basic signature of quantum interferences at the scale of the phase coherence length is strongly reduced in graphene [6, 7]. The physics of proximity induced superconductivity in SN (Superconducting Normal) and SNS structures is also known to be sensitive to quantum interference in the vicinity of the SN interfaces. In the case of graphene, it has been shown theoretically that the Andreev reflection of an electron into a hole at the NS interface, which usually is a retro-reflection, can be specular with a high probability in undoped samples where the Fermi energy lies in the vicinity of the Dirac point [8, 9]. This is predicted to lead to an unusual bias dependence of the differential conductance of the SN or SIN interface [10, 11] as well as multiple Andreev reflections in S/Graphene/S junctions[12].

In this report we show that it is possible to explore this physics by connecting graphene samples to superconducting electrodes and investigate non linear transport at energy scales of the order of and below the superconducting gap.

## II Device Fabrication.

For device fabrication we use the technique of mechanical cleavage (repeated peeling) of highly oriented pyrolytic graphite. This process is performed on an insulating layer of $Al_2O_3$ on a silicon substrate. Few Layer Graphene films (FLG) are spotted as a quasi-transparent film, both with optical and electron microscope observation, compared with thicker graphite films which are not transparent (Fig.1).

In the case of electron microscopy, only the edges of the film can be slightly seen. With this technique, it was possible to obtain samples of a few microns wide. For contacts we use tungsten (W) wires (1 μm x1 μm thick, 100 μm long) grown by decomposition of a metallo-organic vapor (tungsten hexacarbonil) under a Ga Focus Ion Beam (FIB) with a diameter of about 5 nm and accelerating voltage 15 kV. The minimum distance between the W electrodes can be varied between 0.5 and a few microns. The advantage of this technique, compare to the deposition of electrodes using standard lithography techniques, is that it does not involve any deposition of an organic resist on the graphene layers. The tungsten wires grown by FIB at Ga ion current of the order of or above 10 pA, show superconducting characteristics with a critical temperature of 4.5 K and a critical magnetic field, $H_c$, higher than 5 T at 1 K. These parameters are remarkably very reproducible for all wires grown under the same conditions and independent of their geometry. $T_c$ is that of amorphous tungsten [13] but $H_c$ is higher because of a large concentration of impurities. Auger analysis has shown that FIB-deposited tungsten contains about 10% Ga, 10% C, and 5% O [14]. The normal state resistivity of the wires is of the order of few μΩm, which corresponds to an elastic mean free path of the order of 10 nm and a superconducting coherence length of the order of 20-30 nm. We have used previously suspended W electrodes grown by the same technique to investigate proximity induced superconductivity in metallo-fullerenes molecules [15].

The exact number of layers of the sample measured in transport and shown in Fig.1 was difficult to determine due to contamination by insulating amorphous carbon clusters doped with Ga deposited during the focused ion beam (FIB) [16]. However, the thickness of samples with similar optical characteristics to the one measured, was investigated with atomic force microscopy and Raman Spectroscopy like in Ref. [17]. With both characterization techniques the thickness was estimated to be less than 2 nm which corresponds to seven layers or less.

**III Transport Measurements**

Two terminal transport measurements were performed via filtered lines in a dilution refrigerator of base temperature 60 mK. The sample was in most cases current biased with an ac current (frequency 37

Hz) of the order of 0.1 nA superimposed to a dc component for differential resistance measurements. We now present data obtained on a W/FLG/W junction with the minimum distance between the W electrodes equal to 2.5 μm. In the insert of Fig. 2 the temperature dependence of the linear resistance is plotted. The linear resistance measured at room temperature is 1.8 kΩ and increases logarithmically up to 5.2 kΩ when the temperature is lowered down to 1.8 K. This value is very close to the maximum value of resistivity observed in undoped graphene by other groups. Proximity induced superconductivity shows up below $T_c^*=1.7$ K as indicated in the main panel of Fig. 2, via the decrease of the junction resistance by a factor of two but does not reach zero at low temperature.

We now turn to the non linear transport. We first concentrate on $T>T_c^*$ where there is no sign of proximity effect (Fig 3 (a)). A non-linearity in the current-voltage characteristics is observed up to 150 K as shown in Fig.3 (b), where the bias dependence of the differential conductance dI/dV at different temperatures is plotted. It exhibits a minimum at zero bias and a triangular shape which becomes sharper when the temperature is lowered. This behaviour can be related to the characteristic linear dependence of the density of states and indicates that the transmission of the W electrodes is not perfect with the main fraction of the voltage drop through the S/FLG/S junction occurring at the contacts in a symmetrical way. The relatively low transparency is probably due to the etching of the graphene foil below and in the vicinity of the contacts by the FIB. One can also infer from this data and its temperature dependence that the Fermi energy of the Graphene plane is only at few meV away from the Dirac point. We cannot however exclude inhomogeneous doping of the FLG in the vicinity of the electrodes.

Below 1 K, the differential conductance exhibits sharp features at voltages below $2\Delta = 1.6$ meV, where $\Delta=1.76\,k_BT_c$ is the BCS value of the superconducting gap of the W wires with $T_c=4$ K. Since the W electrodes themselves present resistance anomalies above 1mA (far beyond the maximum bias current used in these experiments), we can safely attribute the sub-gap features to the proximity effect through the FLG. Fig. 4 depicts the differential resistance at low temperature in the low bias region. A sharp dip is observed at zero bias but no Josephson current could be detected. This is probably due to the rather large distance between the W electrodes compared to the phase coherence length in graphene which has been found in other experiments [9] not to exceed 1 μm. It is also possible that the absence of supercurrent could be due to specular reflections suggested in Ref. [6] for undoped graphene, since such Andreev pairs having a finite transverse momentum (i.e. along the NS interface) would undergo dephasing as they propagate through the graphene. At finite bias we observe a series of peaks. The position of Two of them can be identified as $2\Delta/e$ and $2\Delta/2e$ (see Fig.4) These can be interpreted as the signature of subgap multiple Andreev reflections [18]. We note the unusually large intensity of the peak at $2\Delta/2e$ which is moreover split, compared to the peak at $2\Delta/e$. The intensity of these peaks are reduced with increasing temperature up to 1 K and their positions are shifted to lower bias with a faster temperature decay than the temperature dependence of the gap of the electrodes as expected from BCS theory which would be nearly independent of Temperature below 2 K. At this stage we cannot exclude [19] the existence of a reduced gap $\Delta' = 0.45$ meV slightly above $\Delta/2$ induced in the FLG in the vicinity of the W electrodes where the doping is the strongest. This could explain a more pronounced anomaly in the differential resistance at $2\Delta'$ compared to the smoother peak observed at $2\Delta$ and the apparent splitting of the peak at $\Delta \sim 2\Delta'$ (see Figs. 4 and 5) . The peaks observed at lower bias can also be related to submultiple values of $2\Delta'$, that is $2\Delta'/2$ and $2\Delta'/3$.

We have also investigated the effect of a magnetic field perpendicular to the plane of the samples. As shown in Fig. 5, the field dependence of the peaks takes place on the Tesla scale, they all shift to lower energy with increasing H, with a stronger variation for the highest energy peak which merges with the second peak at H=2 T. Since any interference effect on the micron length scale would yield a field dependence on the 1mT scale, the observed high field scale confirms the incoherent nature of the transport of Andreev pairs through the FLG junction. The length L of the junction therefore probably verifies: $L_{2\Delta} < L < L_\phi$, where $L_{2\Delta}$ is the inelastic scattering length at the energy $2\Delta$ and $L_\phi$ is the phase coherence length. Since the critical field of the W wire is much larger than 5 T at 60 mK (the maximum field available in this experiment) a possible explanation of the field variations of the peak characteristic energies in our experimental results could be the defocusing of Andreev pair trajectories by magnetic field such that the cyclotron radius corresponds to the distance between the electrodes.

**IV Conclusion and Perspectives.**

We have shown evidence of proximity induced superconductivity in an undoped few-layer-graphene foil connected to superconducting tungsten electrodes. Besides the drop of resistance by a factor of two at low temperature we have observed non-linearities in the differential resistance which at high temperature can be related to the band structure of graphene. In addition, the low temperature differential resistance exhibit peaks at sub-multiple values of double the superconducting gap of tungsten electrodes. These are understood as the signature of incoherent multiple Andreev reflections in the S/Graphene/S junction. Experiments seem also to indicate the existence of another gap smaller than the gap of the W wires probably induced in a small doped region in the very vicinity of the contact region with the electrodes . A systematic investigation of this effect for various distances between the W electrodes and different values of doping is in progress for a deeper understanding of the physics of conversion of Cooper pairs into electron hole pairs in graphene.

**Acknowledgements**

We acknowledge fruitful discussions with F. Guinea and V. Falko. We have also benefited from the expertise of M. Cazayous concerning Raman spectroscopy on our FLG samples. We acknowledge financial support from the FET European network HYSWITCH, and from the ACN program of the French ministry of research.

REFERENCES

[1] P.R.Wallace, Phys.Rev 71, 622 (1947).

[2] A. Castro Neto, F. Guinea, and N. Peres, Phys.Rev. B 73, 205408 (2006).

[3] T. A. Land et al., Surface Sience 264, 261 (1992).

[4] K. S. Novoselov et al., Nature (London) 438, 197 (2005).

[5] Y. Zhang et al., Nature (London) 438, 201 (2005).

[6] S. Morozov et al., Phys. Rev. Lett. 97, 016801 (2006).

[7] E. McCann et al Phys. Rev. Lett. 97, 146805 (2006).

[8] W. J. Beenakker, Phys. Rev. Lett. 96, 246802 (2006).

[9] M. Titov and C. W. J. Beenakker, Phys. Rev. B 74, 041401(R) (2006).

[10] M. Titov, A. Ossipov and C. W. J. Beenakker, cond-mat/0609623 (2006).

[11] Subhro Bhattacharjee and K. Sengupta, cond-mat/0607429 (2006).

[12] J. C. Cuevas and A. Levy Yeyati Phys. Rev. B **74**, 180501 (2006).

[13] N. N. Gribov et al., Physica B 218, 101 (1996).

[14] A. J. DeMarco, J. Melngailis, J. Vac. Sci. Tech. B 19, 2543 5 (2001).

[15] A. Yu. Kasumov et al., Phys. Rev. B 72, 033414 (R) (2005).

[16] The insulating character on the contamination layer was checked by depositing similar electrodes on parts of the substrate without any FLG film.

[17] A. C. Ferrari et al., Phys. Rev. Lett. 97, 187401 (2006).

[18] M. Octavio, M. Tinkham, G. E. Blonder and T. M. Klapwijk, Phys. Rev. B 27, 6739 (1983).

[19] V.Falko (private communication).

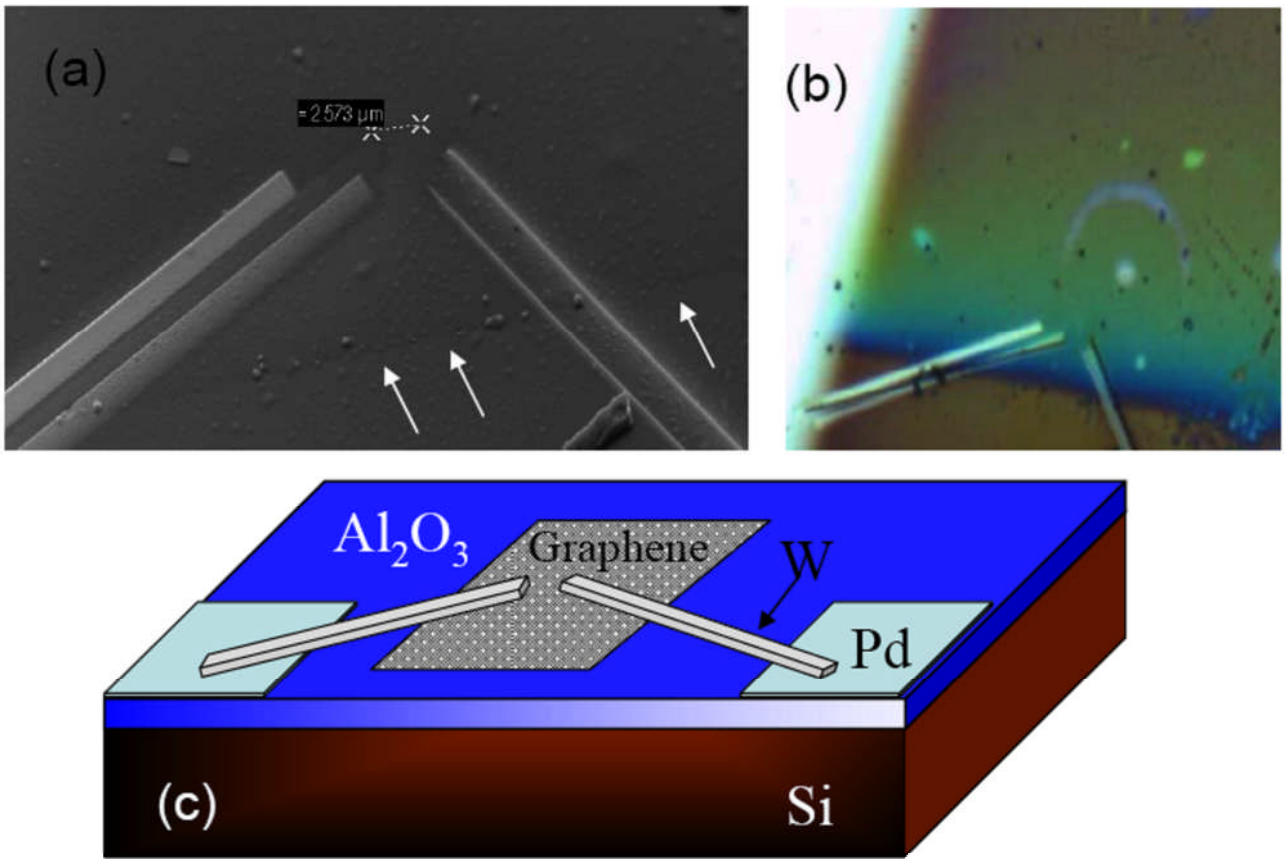

**FIG. 1**

Electron micrograph (**a**) and Optical (**b**) of FLG samples connected to superconducting W Electrodes.

The arrows in (**a**) indicate the edge of the FLG film better seen on the optical image. Only the two inner electrodes on each side of the sample were operational.

(**c**), schematic diagram of the different components of the device.

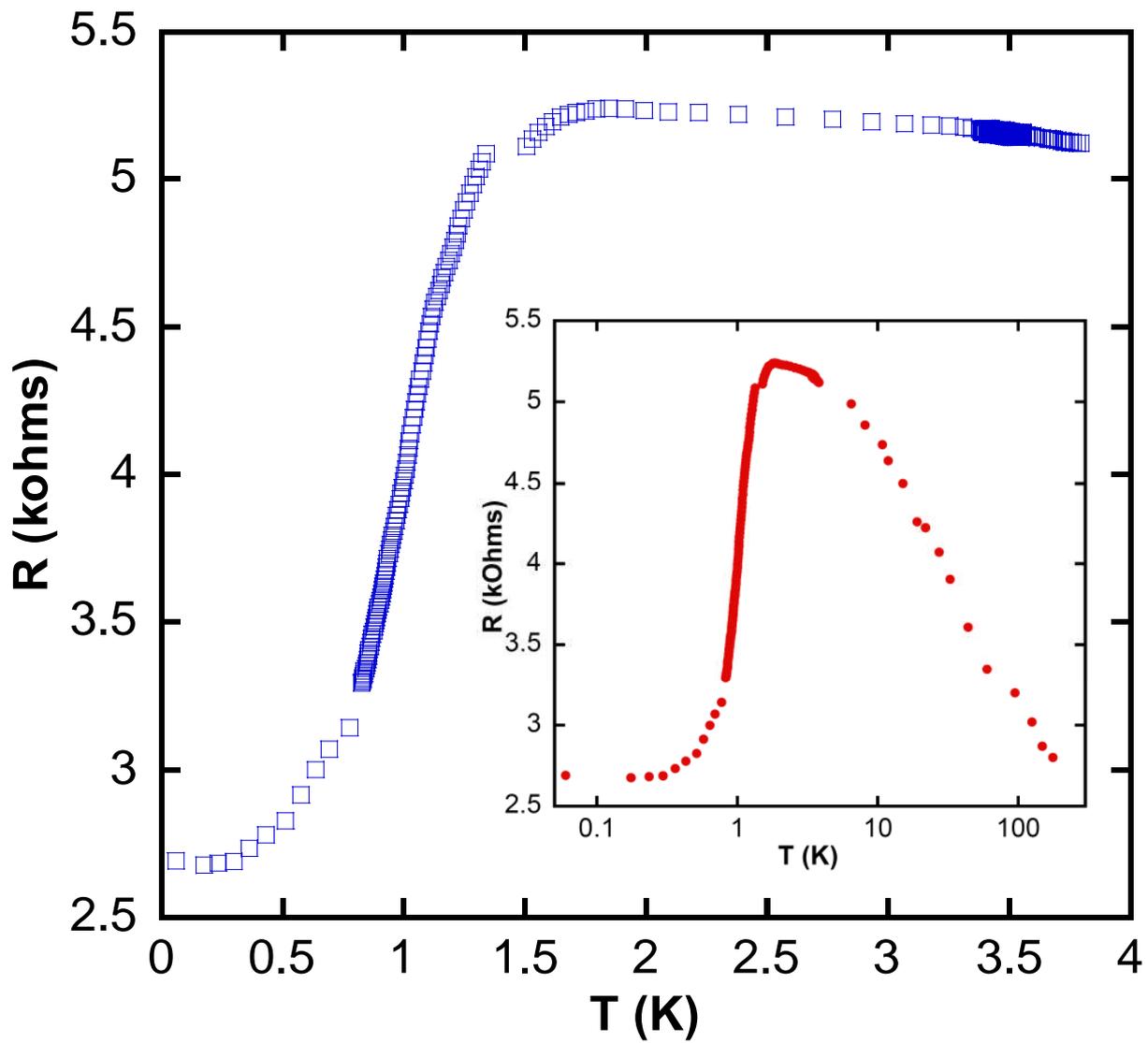

**FIG. 2**

Temperature dependence of the resistance of a W/FLG/W junction with the W junctions distant by 2.5 microns. Main panel shows the transition of the junction which occurs at 1.7 K. Insert: logarithmic dependence of the resistance in the temperature range 200 to 4.2 K

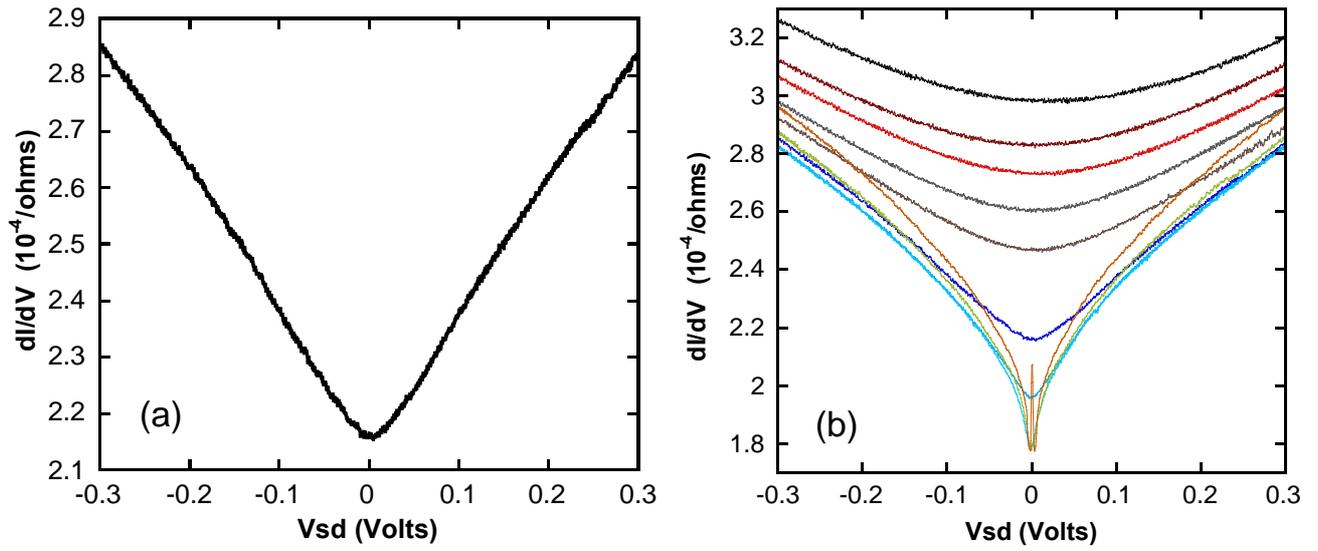

**FIG. 3**

Bias dependence of the differential conductance. The linear increase of the conductance is a characteristic of the linear dispersion relation band structure of graphene. (**a**), Bias dependence at 19 K. (**b**), Bias dependence at different temperatures in the range of 250 mK to 148 K (0.25, 2.5, 3, 4.2, 19, 44, 60, 95, 125, 148 K)

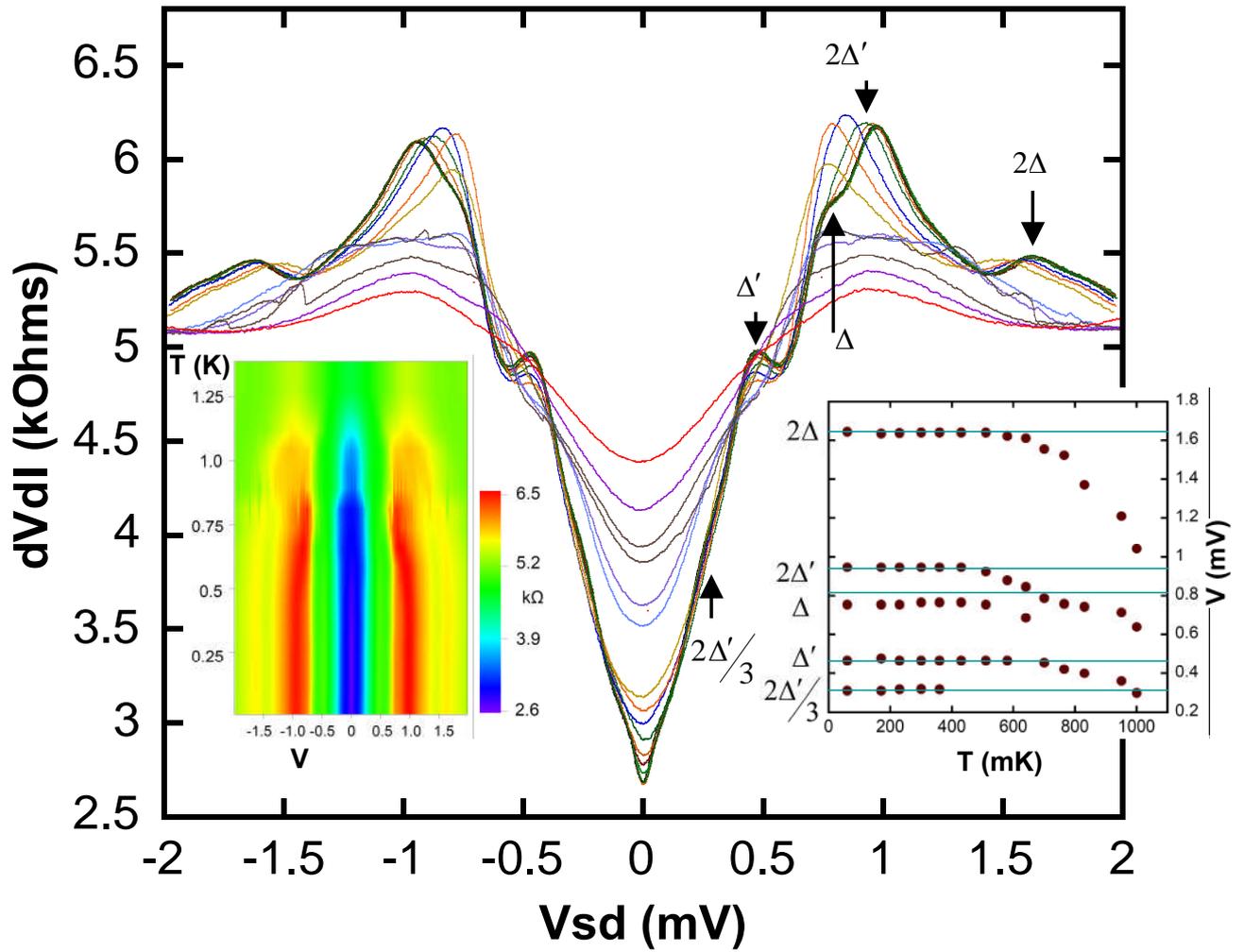

**FIG 4**

Temperature and low bias dependence of the differential resistance showing the presence of peaks at submultiple values of 2Δ and possible 2Δ'. Each trace corresponds to a sweep of the low bias at different temperature between 60 and 1200 mK, with the top red curve corresponding to 1200 mK. Inserts: Left, color plot of the main panel, indicating more clearly the temperature dependence of the MAR peaks shift. Right, temperature dependence of the MAR at submultiple values of 2Δ and possible 2Δ' as shown in main panel.

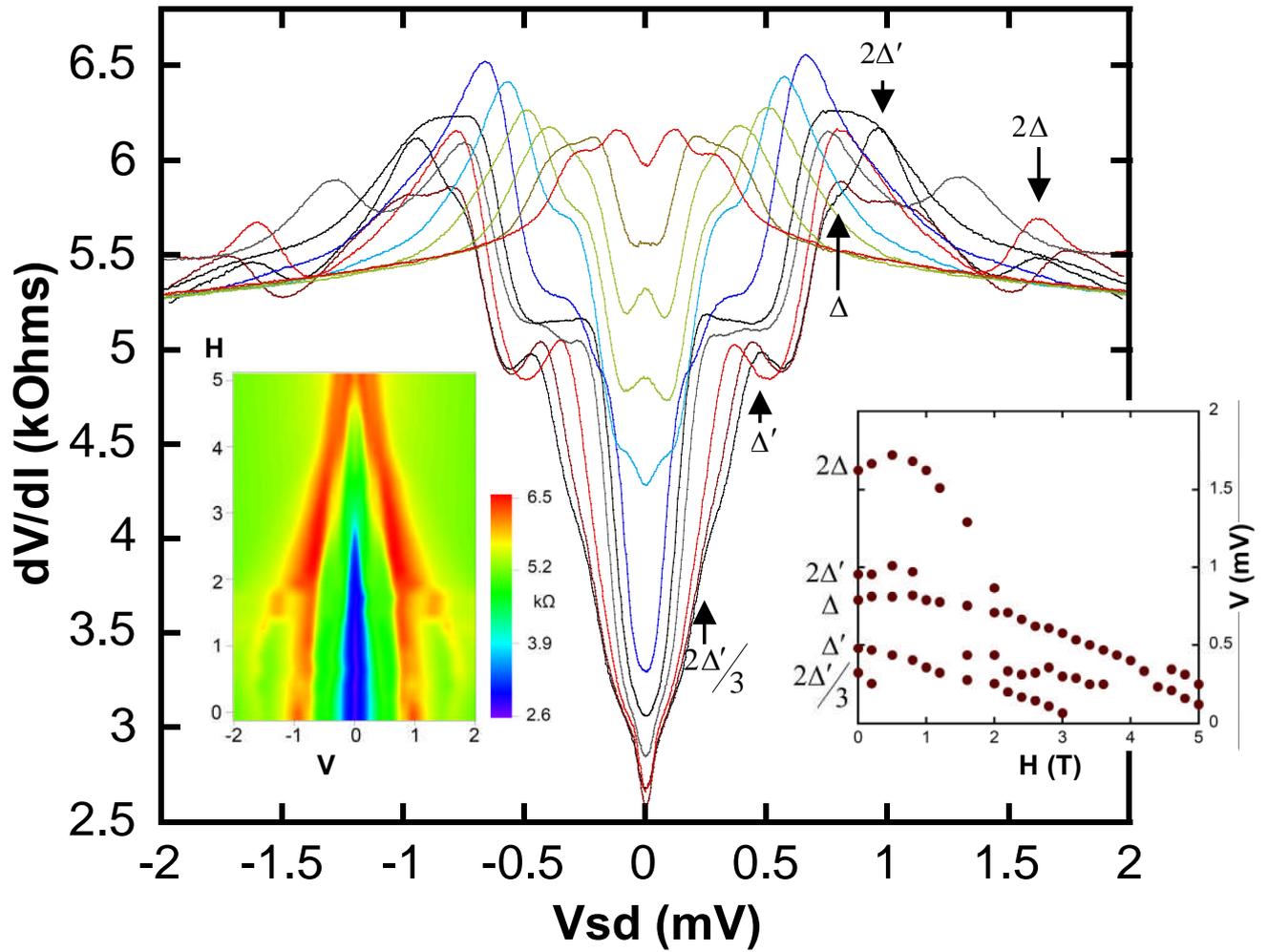

**FIG 5**

Magnetic field and low bias dependence of the differential resistance from 0 to 5 T. Each trace corresponds to a sweep of the low bias at every 0.5 T, with the top red curve corresponding to 5 T. Inserts: Left, color plot of the main panel, indicating more clearly the field dependence of the MAR peaks shift. Right, magnetic field dependence of the MAR at submultiple values of 2Δ and possible 2Δ' as shown in main panel.